\documentstyle[twocolumn,aps,prl,epsf,floats]{revtex}

\begin{document}
\draft
\title{Spin susceptibility and the $\pi$-excitation in
       underdoped cuprates.}
\author{Jan Brinckmann and Patrick A.\ Lee}
\address{Massachusetts Institute of Technology, Cambridge MA\,02139}
\address{ \rm ( 7 October 1997 ) }
\address{ \parbox{14.5cm}{ \rm\small \bigskip
The dynamical spin susceptibility $\chi_\pi''$ at wave vector $(\pi,\pi)$
and the spectrum $\pi''$ of the spin-triplet particle--particle excitation
with center of mass momentum $(\pi,\pi)$ ($\pi$-excitation) are
considered in the slave-boson formulation of the t--J-model. 
Propagators are calculated in a diagrammatic t-matrix approximation in
the d-wave superconducting state for a wide doping range. 
The resulting spectra $\chi_\pi''$ and $\pi''$ both show a resonance at a
doping dependent energy, in qualitative agreement with recent numerical
cluster calculations. In underdoped systems, the peak position is
comparable to that found in neutron scattering experiments. The peak
in $\chi_\pi''$ as well as $\pi''$ is at low doping entirely caused by
spin fluctuations, whereas the triplet particle--particle
channel does not contribute as a collective mode. 
  }}
\maketitle
%
%
%
\newlength{\mysize} 
\def\loadepsfig#1{
 \def\figname{#1}
 \vbox to 10pt {\ }
 \vbox{ \hbox to \hsize {
   \mysize\hsize    \advance \mysize by -20pt 
   \def\epsfsize##1##2{\ifdim##1>\mysize\mysize\else##1\fi}
   \hfill \epsffile{\figname.eps} \hfill 
        } }
 \vbox to 7pt {\ }
 }

The spin-triplet particle--particle excitation (`$\pi$-excitation')
\begin{displaymath}
 \displaystyle \widehat{\pi}^\dagger =
   \sum_k (\cos(k_x) - \cos(k_y) )
   c^\dagger_{-k\uparrow} c^\dagger_{k + q\,\uparrow}
\end{displaymath}
at wave vector $q=(\pi,\pi)$ has been introduced in
\cite{dem95} as a possible explanation for the `41\,meV
resonance' observed in neutron scattering on cuprates in the
superconducting state (see e.g.\ \cite{fon95,bou96}). It has been
argued that $\widehat{\pi}^\dagger$ is an approximate collective
eigenmode of the t--J or Hubbard model. The coupling of 
spin-triplet particle--particle excited states and spin-singlet
particle--hole states in the superconducting phase then
should lead to a resonance in the susceptibility
$\chi_\pi''(\omega)$ at $q=(\pi,\pi)$ at the energy $\omega_0$ of this
$\pi$-mode. 

We compare the susceptibility and the propagator of the $\pi$-excitation
in a slave-boson theory for a wide 
range of hole concentrations (doping). Both calculated spectra
$\chi_\pi''$ and $\pi''$ show a pronounced resonance at the same
energy $\omega_0$\,, which 
is roughly given by the chemical potential $\mu$ as $\omega_0 \approx
2|\mu|$\,. The outcome is in qualitative agreement with the
aforementioned prediction and with recent 
numerical \cite{mei97pre,eder97pre} and diagrammatic calculations
\cite{zha97private}\,. 
However, our interpretation differs from that 
originally envisioned in \cite{dem95}\,: The diagrammatic
spin-triplet particle--particle channel does not contribute as a
collective mode to $\chi_\pi''$ or $\pi''$\,. In underdoped systems 
not far from the transition to the N{\'e}el state, 
the resonance is solely caused by the `RPA channel', which describes spin
fluctuations mediated through spin-singlet particle--hole excitations of
fermions. 

\begin{figure}
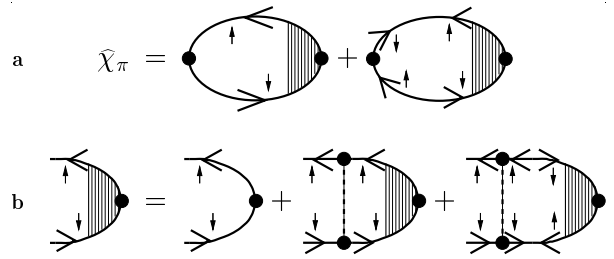

\loadepsfig{figchi}
\caption[\ ]{
  {\bf a:}\,
  Vertex renormalized mean-field susceptibility (part of the 
  t-matrix approximation, see text). 
  {\bf b:}\,
  Vertex function for fermions in the d-wave pairing phase. The dashed
  line represents the nearest neighbor spin and density interaction
  $\sim J$ of the t--J-model. 
  }
 \label{fig-chi}
\end{figure}

\begin{figure}
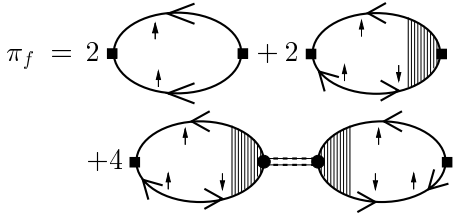

\loadepsfig{figpipi}
\caption[\ ]{
  t-matrix approximation for the $\pi_f$-propagator. The box 
  indicates the phase factor 
  \protect$ \cos(k_x) - \cos(k_y) \protect$\,,
  $k$ is the summed (loop) wave vector.
  The double dashed line stands for the effective
  interaction \protect$ \widetilde{ J}(q,\omega) \protect$ (see text).
  Prefactors count degenerate    exchange parts.
  }
 \label{fig-pipi}
\end{figure}

The `$\pi$-propagator' is given as
\begin{equation}  \label{equ-piprop}
 \displaystyle \pi(\omega)= 
   \langle {\cal T}_\tau \, 
      \widehat{\pi}(\tau) 
      \widehat{\pi}^\dagger(\tau')
   \rangle^\omega \;\,.
\end{equation}
We start from the t--J-model and consider a Gutzwiller-projected
$\pi$-propagator, Eq.\ (\ref{equ-piprop}) with 
$\widehat{\pi} \to P_G \widehat{\pi} P_G$\,. 
The calculations for $\pi(\omega)$ as well as the
susceptibility $\chi_\pi(\omega)$  are performed within the standard 
slave-boson scheme. Diagrammatic expressions are based on a
self-consistent perturbation theory with self energies taken at
Hartree-Fock (mean-field) level. We consider the superconducting state
at very low temperature, i.e., the d-wave pairing phase of fermions
and  fully condensed bosons. The t-matrix approximation for $\chi_\pi(\omega)$
and $\pi(\omega)$ are indicated in Figs.\ \ref{fig-chi} and \ref{fig-pipi}
respectively. 

The susceptibility is given by the
vertex-renormalized mean-field bubbles displayed in Fig.\
\ref{fig-chi}a\,,  which are to be inserted into 
\begin{equation}  \label{equ-rpa}
 \chi_\pi(\omega) = 
  \widehat{\chi}_\pi(\omega) /
  [ 1 - 2J \widehat{\chi}_\pi(\omega) ] \;\,.
\end{equation}
Eq.(\ref{equ-rpa}) represents the particle--hole RPA channel (random
phase approximation). 

The single and double arrowed lines in Fig.\ \ref{fig-chi} stand for
the normal and pairing 
Green's functions of auxiliary fermions. The dashed line in Fig.\
\ref{fig-chi}b is the t--J-model's spin and density interaction for
fermions on two nearest neighbor lattice sites $i,j$\,,
\begin{equation}  \label{equ-inter}
  J \sum_{<i,j>}[ {\boldmath S}_i {\boldmath S}_j - \frac{1}{4}n_i n_j ]
\end{equation}
with $n_i = \sum_\sigma f^\dagger_\sigma f_\sigma$\,. 

The vertex corrections entering $\widehat{\chi}_\pi$ consist of
the spin-singlet particle-hole ($ph$) ladder diagrams shown in Fig.\
\ref{fig-chi}b\,. The double arrowed (anomalous) Green's function
introduces the $ph$ channel in both time directions. In general it
also allows for a coupling of the spin-triplet particle-particle
($pp$) channel into the singlet $ph$ correlation function $\chi_\pi$ by
transforming e.g.\ a spin-up fermion into a spin-down hole and vice versa.
However, the $pp$ channel would appear in $\chi_\pi$ as a vertex-function, 
involving at least one interaction vertex Eq.\ (\ref{equ-inter}) with
equal spin on both sites, which is zero \cite{basand97pre}\,. 
This reflects the fact that Pauli's principle blocks any 
exchange process $\sim J=4 t^2 / U$ for particles with equal spin in
the Hubbard model. Thus the $pp$ channel contributes no spectral
weight to $\chi_\pi$\,. This also holds if the $pp$ channel is
`artificially' switched on by replacing $n_i n_j$ in
Eq.(\ref{equ-inter}) with $g\,n_i n_j$ and turning $g=1 \to g=0$\,:
Recent numerical cluster calculations for the t--J-model
\cite{eder97pre} show that $\chi''$ is not affected by varying the
coefficient of the density--density interaction. In the following we
stick to the case $g=1$\,. 

Numerical calculations in the t-matrix approximation are performed
with mean-field parameters set to reflect a fermion bandwidth of $4 J$ 
and a superconducting gap $\Delta_0 = 40\mbox{\,meV} \approx 0.3
J$\,. As has already been observed in earlier RPA
calculations  \cite{fuk95}\,, an instability to the N{\'e}el state occurs at
an unphysically high hole concentration (doping) $x_c$\,. Since the vertex
corrections of the t-matrix approximation turn out to have no significant
effect, we assume a further renormalization of $J \to \alpha J$ in
Eq.(\ref{equ-rpa})\,. We have chosen $\alpha = 0.5$ such that $x_c$ is
reduced to $\approx 0.02$\,. 

\begin{figure}
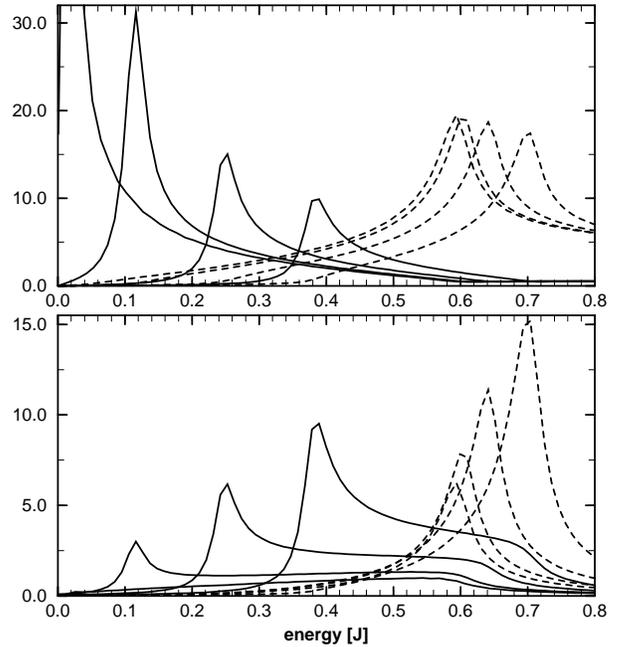
 
\loadepsfig{figtmat}
\caption[\ ]{
  {\bf Top:}\,
  Susceptibility $\chi_\pi''$ in units of $(g\mu_B)^2/(2 J)$ in the
  superconducting 
  phase ($\Delta_0=0.3$) for hole concentrations $x= 0.02, 0.05, 
  0.1, 0.15$ fom left to right. Dashed curves are 
  calculated from the bubble diagrams shown in Fig.\ \ref{fig-chi}a
  (multiplied by $10$). 
  Continuous curves result from the full 
  t-matrix approximation, e.g., are calculated with the
  renormalization through the spin-fluctuation (RPA) channel
  Eq.(\ref{equ-rpa}) taken into account.
  {\bf Bottom:}\,
  $\pi_f$-propagator in units of $1/J$ for the same set
  of $x$ from left to right. Dashed lines result from the
  1st and 2nd (bubble) diagram in Fig.\ \ref{fig-pipi}\,. Continuous
  lines include the RPA channel (3rd diagram).
  }
 \label{fig-tmat}
\end{figure}

Results for $\chi_\pi''(\omega)$ are shown in
Fig.\ \ref{fig-tmat}\,(top) as continuous curves for several hole
densities in the underdoped regime. The dominant feature is
apparently a sharp and strongly doping dependent resonance. Its
position $\omega_0$ shifts from $\approx 0$ at the magnetic
instability ($x= x_c= 0.02$) to higher energies with increased doping,
crossing the anticipated value $40\mbox{\,meV} \approx
0.3J$ around $x= x_m = 0.12$\,. $\omega_0$ is for $x>x_c$ roughly
given by the chemical potential $\mu$ as $\approx 2|\mu|$\,.
This shift of the resonance with hole
concentration is found in neutron scattering experiments on
optimal \cite{fon95,bou96} and underdoped YBCO-compounds
\cite{dai96,fon97}\,. The `optimal' doping 
$x_m\approx 0.12$ found here also compares to experimental
values. However, the spectral weight $\int d^2q\,\chi''(q,\omega) /
\int d^2q$ comes out too small with respect to the experiment
\cite{bou97pre}\,. 

The resonance is caused by the spin-fluctuation RPA channel
Eq.(\ref{equ-rpa})\,:
For comparison, dashed curves in Fig.\ \ref{fig-tmat} (top) show the
results for $\widehat{\chi}_\pi''(\omega)$\,, i.e., the renormalized
bubble diagrams in Fig.\ \ref{fig-chi}a\,. The
position of the doping dependent peak here is bound from 
below by $\approx 2\Delta_0 = 0.6J$\,, even for lowest doping. 
The contribution from the $ph$ vertex corrections is quite
small, $\widehat{\chi}_\pi''$ differs only
slightly from the well known mean-field susceptibility. 

\begin{figure}
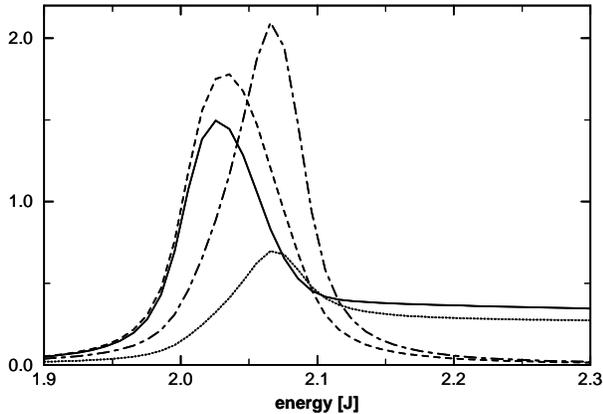
 
\loadepsfig{fighdope}
\caption[\ ]{
  Set of curves for a large chemical potential $\mu=-J$ in the
  superconducting state (see text). $\chi_\pi''$ and $\pi_f''$ (scaled
  $\times 1/4$) are shown
  with the RPA-channel taken into account (continuous and
  dashed line respectively) and with the RPA channel omitted (dotted
  and dashed-dotted line resp.)\,. 
  }
 \label{fig-hdope}
\end{figure}

The results for the susceptibility may be compared to the
$\pi$-propagator Eq.\ (\ref{equ-piprop})\,. In slave-particle 
formulation, with bosons completely condensed at very low temperature,
it reads 
$\displaystyle \pi(\omega) = x^2\, \pi_f(\omega)$\,.
The prefactor $x^2$ is the (mean-field) probability of finding two
empty lattice sites when adding a spin-triplet pair of particles. 
The $\pi$-propagator for fermions $\pi_f(\omega)$ appearing here is
formally identical to Eq.\ (\ref{equ-piprop})\,, its t-matrix
approximation is displayed in Fig.\ \ref{fig-pipi}\,. According to the
discussion given above, the triplet $pp$ channel may contribute only as the
mean-field bubble (1st diagram). The singlet $ph$ channel appears in vertex
renormalizations in the 2nd and 3rd diagram. The 3rd diagram contains
the contribution from the RPA channel via
$\widetilde{ J}(q,\omega) / 2 = 
   \frac{1}{2} J(q) + \frac{1}{4}J(q) \chi(q,\omega) J(q)$\,,
indicated as a double dashed line in Fig.\ \ref{fig-pipi}\,. 

The resulting spectrum $\pi_f''(\omega)$ in the underdoped regime is 
shown in Fig.\ \ref{fig-tmat} (bottom) for the same set of parameters
and hole densities $x$ as the susceptibility. Continuous lines
correspond to the t-matrix approximation, dashed lines are calculated
with the 3rd diagram in Fig.\ \ref{fig-pipi} (the coupling
to the RPA channel) ignored. Again, the effect of the vertex corrections is
negligible, the dashed curves differ only slightly from the mean-field
theory (given by the 1st diagram in Fig.\ \ref{fig-pipi}). Apparently $\pi''$
shows a pronounced peak which occurs at exactly the same position as
the resonance in $\chi_\pi''$\,, if the same approximation is
used for both quantities. As has been pointed out, the
spin-fluctuation (RPA) channel has 
to be taken into account in the underdoped regime $|\mu| <
\Delta_0$\,, where the system is not far from the instability to the
N{\'e}el state, and the RPA dominates $\chi_\pi''$\,. In this case the
peak in $\pi''$ is entirely caused by the 
coupling to spin fluctuations through $\widetilde{ J}$\,. 
Note that its spectral weight decreases with reduced $x$\,, and
vanishes at the transition to the N{\'e}el state ($x=x_c\approx 0.02$)\,. 

The picture changes in a highly overdoped situation $|\mu|\gg\Delta_0$\,: Fig.\
\ref{fig-hdope} shows curves for a large chemical potential
$\mu = -J$ (the breakdown of superconductivity in favor of the fermi-liquid
state $\Delta_0=0$ is ignored for the moment).  The peaks in $\chi_\pi''$ and
$\pi''$ still occur at the same position $\approx 2|\mu|$\,, but the
RPA induces only a slight shift, besides an enhancement of
$\chi_\pi''$\,. In the normal state $\Delta_0=0$ the resonance in
$\chi_\pi''$ vanishes, whereas the peak in $\pi''$ remains as a delta
function ,
$ \pi''(\omega) \sim x^2 \delta(\omega + 2\mu)$\,,
as has also been observed in numerical
calculations on highly doped clusters (referenced in
\cite{zharesponse}). In contrary to the low doping region, the highly
overdoped regime $|\mu|\gg\Delta_0$ is well described by mean-field
theory. 

{\bf Acknowledgments:} Discussions with W.\ Hanke and S.C.\ Zhang are
gratefully acknowledged. One of the authors (JB) acknowledges a
fellowship from the Deutsche Forschungsgemeinschaft, Germany.

\bibliographystyle{prsty}

\end{document}